\documentclass[conference]{IEEEtran}
\IEEEoverridecommandlockouts
\usepackage{cite}
\usepackage{amsmath,amssymb,amsfonts}

\usepackage{algorithmic}
\usepackage{graphicx}
\usepackage{textcomp}
\usepackage{url}
\usepackage{hyperref}
\usepackage{multirow}
\usepackage{multirow}
\usepackage{diagbox}
\usepackage[table]{xcolor}
\begin{document}

\title{Assessing the Robustness of Spectral Clustering for Deep Speaker Diarization}

\author{\IEEEauthorblockN{Nikhil Raghav$^{1,2}$, Md Sahidullah$^{1,3}$}
\IEEEauthorblockA{$^1$Institute for Advancing Intelligence, TCG Centres for Research and Education in Science and Technology, Kolkata-700 091, India \\
$^2$Department of Computer Science, RKMVERI, Howrah-711 202, India\\
$^3$Academy of Scientific and Innovative Research (AcSIR), Ghaziabad-201 002, India\\
e-mail: nikhil.raghav.92@tcgcrest.org, md.sahidullah@tcgcrest.org,}}

\maketitle

\begin{abstract}
Clustering speaker embeddings is crucial in speaker diarization but hasn't received as much focus as other components. Moreover, the robustness of speaker diarization across various datasets hasn't been explored when the development and evaluation data are from different domains. To bridge this gap, this study thoroughly examines spectral clustering for both same-domain and cross-domain speaker diarization. Our extensive experiments on two widely used corpora, AMI and DIHARD, reveal the performance trend of speaker diarization in the presence of domain mismatch. We observe that the performance difference between two different domain conditions can be attributed to the role of spectral clustering. In particular, keeping other modules unchanged, we show that differences in optimal tuning parameters as well as speaker count estimation originates due to the mismatch. This study opens several future directions for speaker diarization research.

\end{abstract}


\section{Introduction} 

Ensuring the accurate automatic annotation of audio recordings based on speaker information is a crucial step for advanced audio analytics. Its application spans from analyzing meeting room data to aiding in courtroom proceedings for advancing AI-assisted judgments. Moreover, it could be valuable in forensic applications, as well as in analyzing conversations between doctors and patients. Over the years, extensive research has been conducted to advance automatic speaker annotation for speech conversations, formally referred to as \emph{speaker diarization} (SD)~\cite{anguera2012speaker}. With the rapid advancements in machine learning, SD systems based on deep learning are demonstrating impressive performance~\cite{park2022review,dawalatabad21_interspeech,wang2024diarizationlm}.

State-of-the-art SD systems yield substantial improvements over classical methods on widely used benchmark datasets including AMI~\cite{carletta2005ami}, CALLHOME~\cite{CallHome}, and VoxConverse~\cite{VoxConverse}. However, SD performance still encounters practical challenges when applied to more realistic conversational speech datasets, including selected subsets within the DIHARD datasets~\cite{ryant2020thirdplan}. Some of the key challenges include the lack of robustness in detecting speech regions, handling speaker overlap, and speaker clustering~\cite{park2022review}. While the first two challenges have received extensive attention in the literature, the problem of speaker clustering has been less explored. This study investigates the robustness of the \emph{spectral clustering} algorithm, which is predominantly utilized in speaker diarization research.

 Traditionally, SD systems have commonly employed \emph{agglomerative hierarchical clustering} (AHC), following a bottom-up approach. In addition to AHC, the SD field has also effectively used basic \emph{k-means}~\cite{wang2018speaker} and \emph{mean-shift clustering methods}~\cite{salmun2017plda}. Nowadays, spectral clustering stands as the default choice~\cite{ning06_interspeech,Park2020AutoTuningSC}, primarily for its simple tuning feature with fewer parameters and for its popularity over other clustering methods. It stems from its flexibility in handling non-linear separations, robustness to noise, ability to capture complex structures, stability in results, and theoretical foundation rooted in \emph{spectral graph theory}~\cite{von2007tutorial,chung1997spectral}.

Additionally, spectral clustering methods are straightforward to apply with deep speaker embeddings~\cite{bai2021speaker}. Here, we explore spectral clustering-based speaker diarization, focusing on its ability to handle dataset mismatches.

The intrinsic and extrinsic variabilities in speech signals cause domain mismatch between training and evaluation data in speech classification tasks~\cite{benzeghiba2007automatic,perkell2014invariance}. Intrinsic variability refers to the natural variations in speech that occur within an individual speaker due to emotion, health, mood, physical condition, etc. On the other hand, extrinsic factors include environmental conditions such as noise and channel effects. The domain mismatch problem is extensively studied in other fields of speech processing, such as \emph{automatic speech recognition}~\cite{deng2013machine}, \emph{speaker recognition}~\cite{bai2021speaker}, \emph{language recognition}~\cite{language_recognition_domain_mismatch,dey2023cross}, \emph{speech enhancement}~\cite{richter2023speech}, \emph{emotion recognition}~\cite{gerczuk2021emonet}, and \emph{audio deepfake detection}~\cite{das2020assessing}. However, for speaker diarization tasks, this issue remains overlooked, likely due to the fact that speaker diarization is principally an unsupervised machine learning problem. Notably, the speaker diarization pipeline~\cite{park2022review} incorporates several data-driven components, making it inherently sensitive to domain-related changes. Given that individual components have been extensively studied for related tasks, this work specifically concentrates on the clustering problem.

\begin{figure*}[t]
    \centering
    \includegraphics[width=8in]{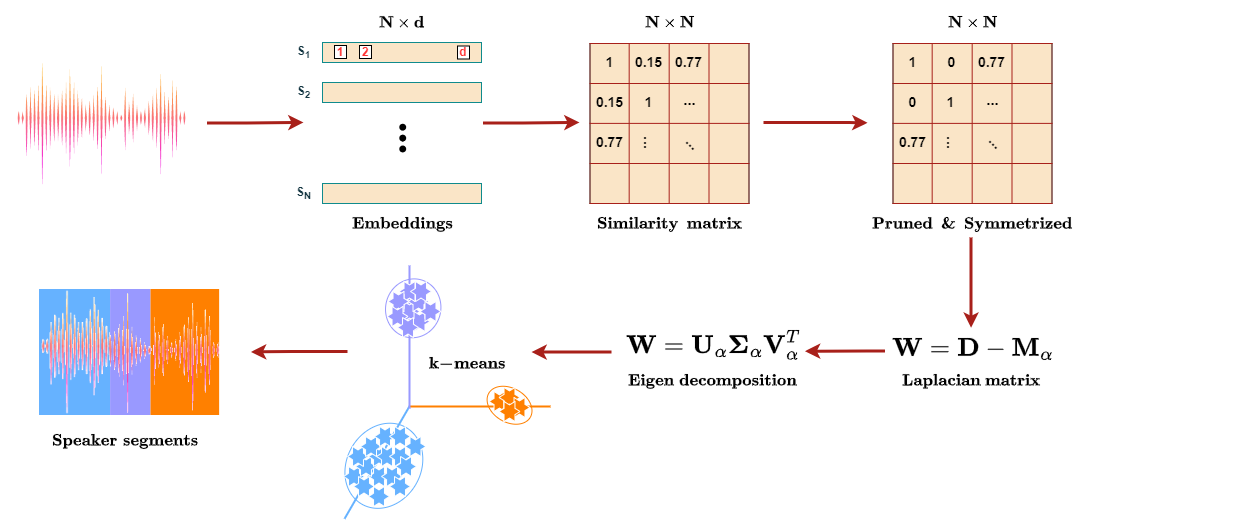}
    \vspace{-0.3cm}
    \caption{Illustration of the steps involved in spectral clustering. The method initiates by extracting speaker embeddings for speech segments. It first calculates the affinity matrix from the $N$ input embeddings situated in a $d$ dimensional space, utilizing cosine similarity as the distance metric. Each row of the affinity matrix prunes smaller values using a tuning parameter $\alpha$. Following symmetrization, it computes an unnormalized Laplacian matrix using the degree matrix $\mathbf{D}$. Subsequently, it applies singular value decomposition (SVD) to the Laplacian matrix $\mathbf{W}$ to derive the leading $k$-eigenvectors. The rows of the eigenvector matrix $\mathbf{U}$ serve as the $k$-dimensional spectral embeddings. Lastly, it performs the standard $k$-means algorithm to cluster these embeddings.}
    \vspace{-0.3cm}
    \label{fig:enter-label}
\end{figure*}

In the present study, we conducted experiments with the AMI~\cite{carletta2005ami} and DIHARD-III~\cite{ryant2020thirdplan} corpora, which consist of varying degrees of domain mismatch. Extending beyond the standard evaluation of the dataset, which typically involves specific development and evaluation sets, we designed a protocol and then performed both same-domain and cross-domain experiments using a state-of-the-art SD system. Furthermore, we have conducted a detailed analysis to uncover the impact of dataset mismatch. Our contributions in this work can be summarized as follows: (i)~We have extensively evaluated the same-domain and cross-domain SD performance for two widely used datasets; (ii)~We have demonstrated how the data mismatch impacts parameter tuning for the clustering problem; (iii)~Our study reveals how the dataset mismatch is related to inherent errors in SD evaluation. To the best of our knowledge, this study is the first to comprehensively investigate the impact of dataset mismatch on SD performance in general as well as spectral clustering in particular.

\section{Speaker diarization system}
\vspace{0.1cm}

\subsection{Review of speaker diarization}
\vspace{-0.1cm}

The SD system typically consists of different interdependent components, namely, \emph{speech enhancement} (SE), \emph{speech activity detection} (SAD), \emph{segmentation}, \emph{speaker embedding extraction}, \emph{clustering}, and \emph{re-segmentation}. The SE module enhances speech by reducing the impact of noises. Speech and non-speech regions are identified by SAD, and speech regions are segmented before being fed into a speaker embedding extractor. Then clustering is performed to group segment-wise speaker embeddings into speaker-wise clusters. Finally, the labels are refined with re-segmentation to address speaker overlap issues, especially in segment boundary regions.

State-of-the-art SD systems either use the components in a \emph{pipelined} framework or are trained in an integrated \emph{end-to-end} fashion. The latter approach requires a single DNN model, which may facilitate easy domain adaptation and offer flexibility. In the former approach, all individual components are separately crafted. Except for segmentation, which typically uses fixed-length chunking, each component is trained for a specific task, and then the given speech data or its representation is processed. This may introduce domain mismatch problems specially when the data used for system development and evaluation are different. We investigate the impact of domain mismatch issue in the context of clustering.

\subsection{Spectral clustering and its application in SD}
\vspace{0.1cm}

Spectral clustering technique is a graph-based approach for clustering that utilize the spectrum (eigenvalues) of the similarity matrix for dimensionality reduction before clustering in lower dimensions. This method finds extensive application, especially in image segmentation~\cite{shi2000normalized,ng2001spectral,von2007tutorial}.

The concept of spectral clustering for SD was first introduced in~\cite{ning2006spectral} and later applied in subsequent works~\cite{bassiou2010speaker,iso2010speaker}. Its utilization has become widespread, particularly with the emergence of speech embeddings~\cite{shum12_interspeech,sell2014speaker,wang2018speaker,aronowitz20_interspeech,lin19_interspeech}. The work reported in~\cite{Park2020AutoTuningSC} developed an auto-tuning framework of spectral clustering. Nevertheless, it suffers from computational inefficiency, rendering standard spectral clustering, as advocated in~\cite{dawalatabad21_interspeech}, a more preferable option. Our method follows this framework (shown in Fig.~\ref{fig:enter-label}), which is summarized below.

Given a finite set $\mathbf{S} =$ \{$s_{1}, s_{2}, ..., s_{N}$\} of $N$ embeddings extracted from their corresponding $N$ speech segments. Our objective is to cluster them into $k$ speaker classes. The steps for spectral clustering can be described as follows:

\begin{itemize}
    \item First, the affinity matrix $\mathbf{M}\in \mathbb{R}^{N\times N}$ is computed. Each entry is calculated using the cosine similarity $s_{i}$ and $s_{j}$ where $i$ and $j$ are the corresponding speech segment indexes.

    \item To alleviate the effect of any unreliable values, higher values in each row of $\mathbf{M}$ is retained while remaining values are made zero. A pruning parameter $\alpha$ is defined and smaller $\left \lceil N(1-\alpha) \right \rceil$ entries are made to zero.

    \item The resulting matrix is symmetrized to create $\mathbf{M}_{\alpha}$, i.e., $\mathbf{M}_{\alpha} = (\mathbf{M}+\mathbf{M}^{\top})/2$.
    
    \item The subsequent step involves the computation of an unnormalized Laplacian matrix $\mathbf{W} = \mathbf{D} - \mathbf{M}_{\alpha} $~\cite{von2007tutorial}, where $\mathbf{D}$ is the degree matrix.

    \item Eigendecomposition is performed on $\mathbf{W}$, to estimate the number of speakers $k$, using the maximum eigengap approach~\cite{von2007tutorial}.

    \item The first $k$ eigenvectors are computed. The rows of the resulting matrix of eigenvectors are the $k$ dimensional spectral embeddings. 

    \item Finally, the standard $k$-means algorithm is used to cluster these embeddings.
\end{itemize}
To estimate the optimal pruning parameter, $\alpha$ is varied from 0 to 1 with a typical step size of $0.01$. The value of $\alpha$ that exhibits the best diarization performance on the tuning set is then utilized for the evaluation set.

The work in~\cite{Park2020AutoTuningSC} proposed using a proxy DER measure to optimize clustering performance on each recording. However, this approach may be unrealistic in real-time scenarios, as it involves the eigenvalue decomposition as many times as the number of segments in a recording. 

\vspace{-0.1cm}
\section{Experimental setup}
\vspace{-0.1cm}

\subsection{Dataset}
\vspace{-0.1cm}
Our current study utilizes two widely used speech corpora: the augmented multi-party interaction (AMI) meeting corpus~\cite{carletta2005ami} and DIHARD III~\cite{ryant21_interspeech}. 

\subsubsection{AMI corpus} 
The AMI is a multi-modal dataset consisting of 100 hours of meeting recordings~\cite{carletta2005ami}. We have used the speech portion. Several types of microphones, including close-talking and far-field, were used to collect the speech data from different sites located in meeting rooms at the University of Edinburgh (U.K), IDIAP (Switzerland), and TNO Human Factors Research Institute (The Netherlands). We performed the experiments using the official \emph{Full ASR Corpus} split\footnote{\url{https://groups.inf.ed.ac.uk/ami/corpus/datasets.shtml}}, with TNO meetings excluded from the development (Dev) and evaluation (Eval) sets. Finally, it consists of 14 recordings in the Dev and 12 recordings in the Eval set. The recordings are broadly categorized into three categories representing three different room environments and scenarios. Some of the recordings share the same room and speakers but were recorded in a different session. For this study, we have used three different microphone configurations: Mix-Headset, Mix-Lappel, Mic-Array (first channel of microphone array Array1) as different domains.

\subsubsection{DIHARD-III corpus}
The DIHARD-III corpus was created during the third DIHARD speaker diarization challenge (DIHARD III)~\cite{ryant21_interspeech}. It includes data from 11 diverse domains, each characterized by varying numbers of speakers, conversation type and speech quality. These domains range from relatively high-quality recording scenarios such as broadcast interview and courtroom interactions to low-quality far-field challenging environments like restaurants and webvideo. For this study, we focus on seven domains: \emph{broadcast interview}, \emph{court}, \emph{cts}, \emph{maptask}, \emph{meeting}, \emph{socio lab}, and \emph{webvideo}. We excluded four domains due to either containing single-speaker data (in audiobooks) or personal identifying information (PII) regions (in clinical, restaurant, socio field). The PII regions are processed with low-pass filtering or insertion of tones or zeroing out of samples~\cite{ryant2020thirdplan}, and including them may negatively impact overall clustering performance.

\vspace{-0.1cm}
\subsection{SpeechBrain system}
\vspace{-0.1cm}
We have used SpeechBrain toolkit for our experiments~\cite{speechbrain}. Our implementation is based on the AMI recipe in this toolkit~\footnote{\url{https://github.com/speechbrain/speechbrain/tree/develop/recipes/AMI}}. This uses pre-trained ECAPA-TDNN speaker embedding extractor trained on VoxCeleb dataset\cite{nagrani17_interspeech,chung18b_interspeech}. The extracted embeddings are 192 dimensional calculated from the second last layer of ECAPA-TDNN~\cite{dawalatabad21_interspeech}. We have used ground-truth SAD to avoid the influence of selection of SAD. The segment size is fixed at $3.0$s with $1.5$s of overlap. We have used cosine scoring for similarity measure.

\vspace{-0.1cm}
\subsection{Evaluation metric}
\vspace{-0.1cm}
In evaluating the performance of the SD system, we utilize the \emph{diarization error rate} (DER) as the primary evaluation metric, as outlined in~\cite{10.1007/11965152_28}. To compute this metric, first speaker correspondance between the ground-truth and the predicted speakers is computed using Hungarian algorithm \cite{kuhn1955hungarian}. DER is comprised of three key errors: \emph{missed speech}, \emph{false alarm of speech}, and \emph{speaker error}~\cite{anguera2006robust}. DER is quantified as the ratio of the combined duration of these three errors to the total duration. A forgiveness collar of {\color{green!55!blue}$0.25$s} is kept in the experiments, to mitigate the effect of inconsistencies in the manual ground-truth annotations.

\vspace{-0.1cm}
\section{Results}
\vspace{-0.1cm}

\subsection{Results on AMI corpus}
\vspace{-0.1cm}
\begin{table}[]
\renewcommand{\arraystretch}{1.3}
    \centering
    \caption{SD performance (in terms of \% of DER) on the AMI dataset for different tuning data and test data conditions.}
    \vspace{-0.3cm}
    \begin{tabular}{|c|c|c|c|}
    \hline
    \multirow{2}{*}{Tuning domain}& \multicolumn{3}{|c|}{Evaluation domain}\\
    \cline{2-4}
     & Mix-Headset & Mix-Lapel & Mic-Array  \\
    \hline
    Mix-Headset & \cellcolor{gray!50}\textbf{1.58} & 2.18 &  \textbf{3.56}\\
    Mix-Lapel & 1.78 & \cellcolor{gray!50}2.26 & 3.80 \\
    Mic-Array & 1.77 & \textbf{2.15} & \cellcolor{gray!50}4.33  \\
    \hline
    \end{tabular}
    \label{Table:Results:AMI}
    \vspace{-0.3cm}
\end{table}

\begin{table*}[t]
\renewcommand{\arraystretch}{1.3}
    \centering
    \caption{SD performance (in terms of \% of DER) on the third DIHARD dataset for different tuning data and test data conditions.}
    \vspace{-0.3cm}
    \begin{tabular}{|c|c|c|c|c|c|c|c|}
    \hline
     \multirow{2}{*}{Tuning domain}& \multicolumn{7}{|c|}{Evaluation domain}\\
     \cline{2-8}
     & broadcast interview&court &cts &maptask &meeting &socio lab &webvideo  \\
     \hline
     broadcast interview & \cellcolor{gray!50}3.58 & 3.13 & 6.58 & 1.23 & 17.50 & 1.95 & 35.12 \\
     court & 4.18 & \cellcolor{gray!50}\textbf{2.08} & 7.03 & 4.70 & \textbf{14.97} & 2.00 & 38.50 \\
     cts & 4.02 & 11.29 & \cellcolor{gray!50}\textbf{6.56} & \textbf{0.93} & 24.43 & \textbf{1.82} & \textbf{33.81} \\
     maptask & 4.48 & 11.29 & \textbf{6.56} & \cellcolor{gray!50}\textbf{0.93} & 24.43 & 1.83 & 33.99 \\
     meeting & \textbf{2.54} & 5.29 & \textbf{6.56} & 0.98 & \cellcolor{gray!50}18.97 & 1.92 & 36.79 \\
     socio lab & 4.02 & 9.09 & 6.58 & 0.96 & 24.53 & \cellcolor{gray!50}1.86 & 34.09 \\
     webvideo  & 4.48 & 11.29 & \textbf{6.56} & \textbf{0.93} & 24.43 & \textbf{1.82} & \cellcolor{gray!50}35.84 \\   
    \hline
    \end{tabular}
    \label{Table:Results:DIHARD}
    \vspace{-0.3cm}
\end{table*}
Table~\ref{Table:Results:AMI} compares the same- and cross-domain performance of the SD system. We tune and test the SD system on the three single-channel microphone settings. Table~\ref{Table:Results:AMI} indicates that Mix-Headset shows lowest DER when used for both train and test whereas Mic-Array in both tune and test shows poorest performance by demonstrating highest DER. This table also reveals that the performance obtained with Mix-Headset as tuning data gives lower DERs than other configurations.

\vspace{-0.1cm}
\subsection{Results on DIHARD-III corpus}
\vspace{-0.1cm}
Table~\ref{Table:Results:DIHARD} compares the cross-domain and same-domain performance for the DIHARD-III corpus. From the same-domain results, we observe that broadcast interview, maptask, and socio lab show relatively lower DER compared to other cases, especially meeting and webvideo. The cross-corpora results, on the other hand, reveal several interesting findings. First, we observe that using data from some of the domains better generalizes to other domains. For example, cts in tuning helps to obtain the best DER in four of the seven cases. Second, similar to the results on AMI, the same corpora in tuning and evaluation do not necessarily give the best DER. This contradicts the findings in other speech classification tasks where the matched data in training and testing usually show the best performance. The exception in this SD case might be due to the fact that the datasets from the specific domain might have intra-domain variations, such as the number of speakers. Table~\ref{Table:Results:DIHARD} also shows that some of the domains are more sensitive to the parameter tuning data than others. For example, the standard deviation of DERs in evaluating broadcast interview and cts are 0.67 and 0.18, respectively.

\vspace{-0.1cm}
\subsection{Analysis of the impact of tuning parameter}
\vspace{-0.1cm}

\begin{figure}
    \centering
    \includegraphics[width=3in]{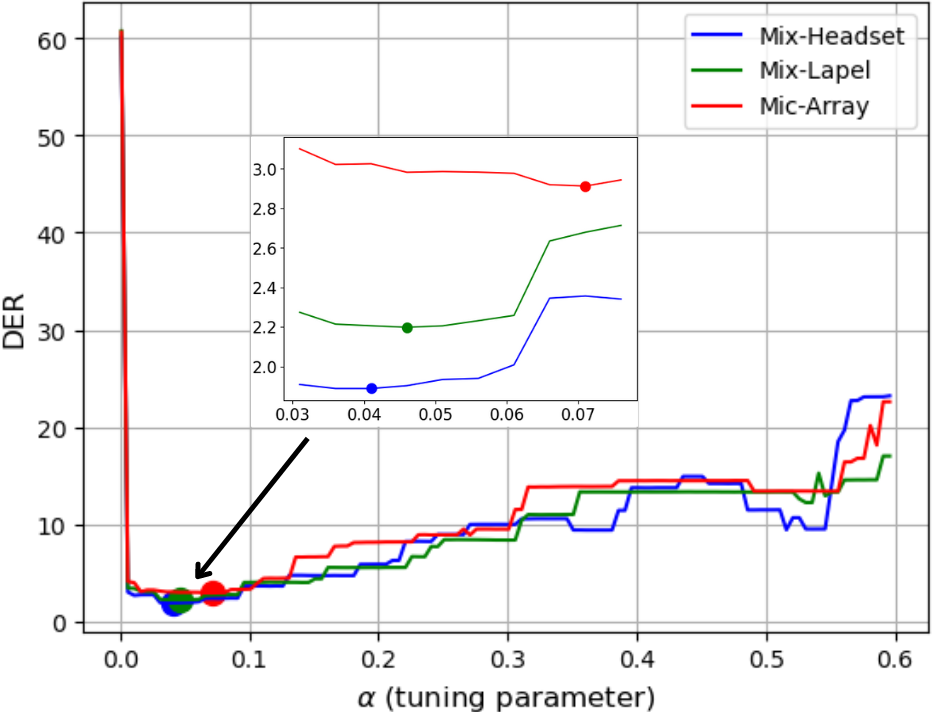}
    \vspace{-0.3cm}
    \caption{Plot showing the impact of the tuning parameter $\alpha$ on DER for the three microphone types in AMI development subsets. The circles represent the lowest DERs corresponding to each condition. [Best view in color.]}
    \vspace{-0.3cm}
    \label{Fig:tuningAMI}
\end{figure}

\begin{figure}
    \centering
    \includegraphics[width=3in]{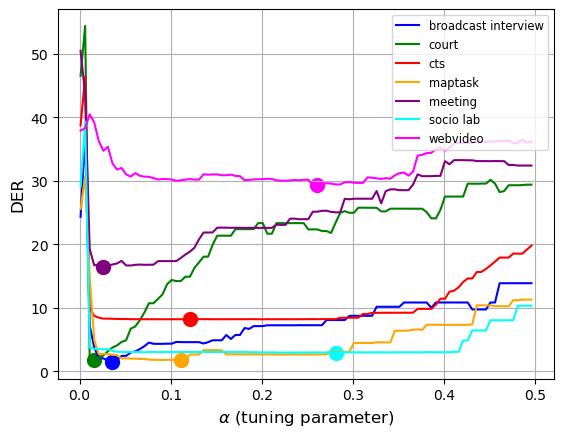}
    \vspace{-0.3cm}
    \caption{Plot showing the impact of the tuning parameter $\alpha$ on DER for the seven domains in DIHARD III development subsets. The circles represent the lowest DERs corresponding to each condition. [Best view in color.]}
    \label{Fig:tuningDIHARD}
    \vspace{-0.3cm}
\end{figure}

The selection of tuning parameters is crucial, and we conduct further experiments to understand the role of the dataset in tuning parameter selection. We conduct same-domain (SD) experiments on development data and analyze the impact of tuning parameters on DER. Figures~\ref{Fig:tuningAMI} and~\ref{Fig:tuningDIHARD} show the results for the different domains of the development set of AMI and DIHARD-III, respectively. We observe that the optimal tuning parameters for AMI are closer in numerical values than those for DIHARD-III. This is due to smaller intra-domain variations in AMI compared to DIHARD-III. For DIHARD-III, we also observe that even when selecting optimal tuning parameters in each subset, we may not improve the SD performance. This demonstrates the limitation of spectral clustering for some difficult domains such as webvideo.

\vspace{-0.1cm}
\subsection{Impact on speaker counting}
\vspace{-0.1cm}

In a subsequent experiment, we evaluated the performance of spectral clustering on speaker count estimation. Table~\ref{SpeakerCount} displays the results of speaker count errors in terms of the average error per audio recording. The results for the AMI corpus indicate that the spectral clustering algorithm accurately estimates the number of speakers for the development set. However, errors are higher for the evaluation set, and they increase with the degradation in speech quality. On the other hand, for the DIHARD-III, we observed noticeable speaker count estimation errors. Our study also reveals that the most challenging domains (as indicated by the DER performance in Table~\ref{Table:Results:DIHARD}), such as meetings and webvideo, exhibit higher errors. This opens up opportunities to revisit the speaker clustering to tackle the speaker count estimation issue.

\begin{table}[h]
\renewcommand{\arraystretch}{1.3}
    \centering
    \caption{Speaker count estimation error, between the ground-truth and the system generated.}
    \vspace{-0.3cm}
    \begin{tabular}{|c|c|c|c|}
    \hline
    \multirow{2}{*}{Dataset}& \multirow{2}{*}{Domain} & \multicolumn{2}{|c|}{Speaker Error}\\
    \cline{3-4}
     & &Dev & Eval \\
    \hline
     & Mix-Headset &0.00 & 0.07 \\
    AMI Corpus & Mix-Lapel&0.00 & 0.21 \\
     & Mic-Array&0.00 & 0.35 \\
     \hline
     & broadcast interview& 0.75 & 0.50  \\
     & court& 0.75 & 0.91  \\
     & cts& 0.08 & 0.11  \\
    DIHARD III & maptask&0.30 & 0.04 \\
    & meeting& 1.71 & 1.00  \\
    & socio lab& 0.69 & 0.75  \\
     & webvideo &2.47 & 2.43\\
    \hline
    \end{tabular}
    \vspace{-0.3cm}
    \label{SpeakerCount}
\end{table}

\vspace{-0.1cm}
\section{Conclusions}
\vspace{-0.1cm}
This work investigates the issues of spectral clustering for SD across multiple domains. We have demonstrated that spectral clustering plays a pivotal role from two aspects. First, it estimates the number of speakers in an efficient way. Then, it assigns the cluster labels by clustering in a subspace. Our experimental results with two popular datasets show that selection and optimization of pruning parameters is an important issue. This task is also challenging when performing SD experiments in a domain mismatch scenario. Our study opens several research directions by revealing this intrinsic issue. For example, investigation can be conducted to develop computationally efficient automatic estimation of pruning parameter in a recording-specific way.

Investigation of this issue with other advanced speaker embeddings could be another future direction.

\section*{Acknowledgment}

The primary author expresses sincere gratitude to the Linguistic Data Consortium (LDC) for the LDC Data Scholarship, which enabled access to the DIHARD-III dataset.

\bibliographystyle{ieeetr}
\bibliography{references}

\end{document}